\documentclass[aps,prb,twocolumn,showpacs,superscriptaddress,preprintnumbersfleqn,floatfix]{revtex4-2}

\usepackage{graphicx}
\usepackage{dcolumn}
\usepackage{bm}
\usepackage{amsmath, amssymb}

\begin{document}

\title{Magnetic-field-induced magnon portfolio in a van der Waals magnet}

\author{T. Riccardi}
\affiliation{Laboratoire National des Champs Magn\'etiques Intenses, CNRS, LNCMI, Universit\'e Grenoble Alpes, Univ Toulouse, INSA Toulouse, EMFL, F-38042 Grenoble, France}
\affiliation{Universit\'e Grenoble Alpes, CNRS, Grenoble INP, Institut N\'eel, 38000 Grenoble, France}

\author{F. Le Mardélé}
\affiliation{Laboratoire National des Champs Magn\'etiques Intenses, CNRS, LNCMI, Universit\'e Grenoble Alpes, Univ Toulouse, INSA Toulouse, EMFL, F-38042 Grenoble, France}
\author{L.A. Veyrat de Lachenal}
\affiliation{Laboratoire National des Champs Magn\'etiques Intenses, CNRS, LNCMI, Universit\'e Grenoble Alpes, Univ Toulouse, INSA Toulouse, EMFL, F-38042 Grenoble, France}
\affiliation{Universit\'e Grenoble Alpes, CNRS, Grenoble INP, Institut N\'eel, 38000 Grenoble, France}

\author{A. Pawbake}
\affiliation{Laboratoire National des Champs Magn\'etiques Intenses, CNRS, LNCMI, Universit\'e Grenoble Alpes, Univ Toulouse, INSA Toulouse, EMFL, F-38042 Grenoble, France}

\author{I. Plutnarova}
\affiliation{Department of Inorganic Chemistry, University of Chemistry and
Technology Prague, Technick\'a 5, 166 28 Prague 6, Czech Republic}
\author{Z. Sofer}
\affiliation{Department of Inorganic Chemistry, University of Chemistry and
Technology Prague, Technick\'a 5, 166 28 Prague 6, Czech Republic}

\author{G. Jacquet}
\affiliation{Aix-Marseille Universit\'e, CNRS, CINaM, 13284 Marseille, France}

\author{F. Petot}
\affiliation{Aix-Marseille Universit\'e, CNRS, CINaM, 13284 Marseille, France}

\author{A. Sa\'ul}
\affiliation{Aix-Marseille Univ, CNRS, CINaM, Marseille, France}

\author{B. Gr\'emaud}
\affiliation{Aix-Marseille Univ, Universit\'e de Toulon, CNRS, CPT, Marseille, France}

\author{A. L. Barra}
\affiliation{Laboratoire National des Champs Magn\'etiques Intenses, CNRS, LNCMI, Universit\'e Grenoble Alpes, Univ Toulouse, INSA Toulouse, EMFL, F-38042 Grenoble, France}

\author{M. Orlita}
\affiliation{Laboratoire National des Champs Magn\'etiques Intenses, CNRS, LNCMI, Universit\'e Grenoble Alpes, Univ Toulouse, INSA Toulouse, EMFL, F-38042 Grenoble, France}
\author{J. Coraux}
\affiliation{Universit\'e Grenoble Alpes, CNRS, Grenoble INP, Institut N\'eel, 38000 Grenoble, France}

\author{C. Faugeras}
\affiliation{Laboratoire National des Champs Magn\'etiques Intenses, CNRS, LNCMI, Universit\'e Grenoble Alpes, Univ Toulouse, INSA Toulouse, EMFL, F-38042 Grenoble, France}
\author{B. A. Piot}\email{benjamin.piot@lncmi.cnrs.fr}
\affiliation{Laboratoire National des Champs Magn\'etiques Intenses, CNRS, LNCMI, Universit\'e Grenoble Alpes, Univ Toulouse, INSA Toulouse, EMFL, F-38042 Grenoble, France}

\date{\today}

\begin{abstract}

Magnonic excitations are investigated in chromium oxychloride (CrOCl), a van der Waal (vdW) antiferromagnet prone to a multitude of magnetic phase transitions, with absorption experiments in a broad continuous energy range. At low magnetic fields, the magnon spectra show a strong bi-axial anisotropy and inform on the relative weights of the effective exchange coupling and the system anisotropies. As the magnetic field increases, magnons characteristic of a canted phase are first observed, with peculiarities attributed to in-plane anisotropies and magnon-magnon coupling. Subsequently, a hysteretic magnon spectrum appears as the system transitions to a ferrimagnetic state, with two new magnon branches partly coexisting with the lower energy canted phase branch, indicating the formation of spatially separated magnetic phases. Further changes in the magnon spectrum in higher magnetic fields accompany transitions to the different canted magnetic phases previously reported. Our experiments show that competing exchange interactions and ground states broaden the options to generate different kinds of magnonic excitations in the same vdW material upon the variation of external parameters.

\end{abstract}
\maketitle

Magnons are the quanta of collective and charge-neutral excitations of spins (spin waves) in magnetic materials. Their dispersion, propagation, and interactions with other crystal excitations provide invaluable information about spin-spin interactions and the coupling of different degrees of freedom in the solid. Additionally, they hold promise for low-loss and high-speed information transfer owing to their chargeless nature and highly-coherent space and time propagation~\cite{Chumak2015}.
In van der Waals (vdW) magnets, where the interlayer exchange coupling is generally weak compared to covalently bond solids, the magnon dynamics of interlayer antiferromagnets (AFM) can be brought down from the THz to the GHz regime, such as in the well-studied  CrCl$_{3}$~\cite{MacNeill2019}, CrPS$_{4}$~\cite{Li2023} or CrSBr~\cite{Cho2023}. This constitutes an ideal platform to extend the study and exploitation of spin dynamics by using microwave/radio frequency  techniques (e.g. including inductive methods). Another possibility to effectively modulate the exchange coupling strength is to consider systems with competing interactions, wherein non-trivial spin orders, and even frustration physics, tend to develop. In such systems, the balance between competing interactions may be modelled via an effective exchange coupling $J_{eff}$, to which the magnons energy is generally proportional in the presence of anisotropy in the system. The effects of the competition can moreover be controlled with external parameters such as magnetic fields~\cite{Wosnitza2016} or hydrostatic pressure~\cite{Thede2014}, resulting in a variety of possible ground states. The widely tunable nature of vdW magnets (see Ref.~\onlinecite{Wang2022} for a review) together with strongly competing interactions should offer multiple options to access to different kinds of magnons, living in a wide range of energies, all that in a single material.

CrOCl, which has been first synthesized in 1974~\cite{Christensen1974} and more deeply studied in its bulk form since the 2010's, is an example of such a vdW magnet. More recently, CrOCl was exfoliated and studied down to a few or even single layer~\cite{Gu2022,Zhang2023} and found to be robust in air, a key asset compared to historically discovered vdW magnets~\cite{Gong2017}. Bulk CrOCl  hosts a low temperature anti-ferromagnetic order at ambient pressure and in absence of magnetic field, which results from the competition of opposite intralayer exchange between Cr spins, and displays a series of magnetic-field-induced ground states~\cite{Reuvekamp2014,Pawbake2025}, the spin excitations of which have so far remained unexplored.

In this article, we report a GHz/THz absorption study of bulk CrOCl over a continuous, broad energy range in the presence of high magnetic fields, in which various magnon branches are identified in the multiple magnetic-field-induced phases. The AFM magnon spectra demonstrate a strong bi-axial anisotropy and set constraints between the frustration-induced effective exchange coupling and the system anisotropies. Magnons characteristic of a canted phase are then observed, with peculiarities attributed to anisotropies and magnon-magnon coupling. As the magnetic field is further increased, a hysteretic magnon spectrum appears as the system transitions to a ferrimagnetic state (FiM), with two magnon branches  partly coexisting with the lower energy canted phase branch indicating the formation of spatially separated magnetic phases. Further changes in the magnon spectra appear to correspond to the different canted phases inferred from previous studies in higher magnetic fields~\cite{Reuvekamp2014,Pawbake2025}. Our work shows that competing exchange interactions and ground states broaden the options to generate different kinds of magnonic excitations in the same vdW material upon the variation of external parameters.

\begin{figure*}[t]
\begin{center}
\includegraphics[width=18cm]{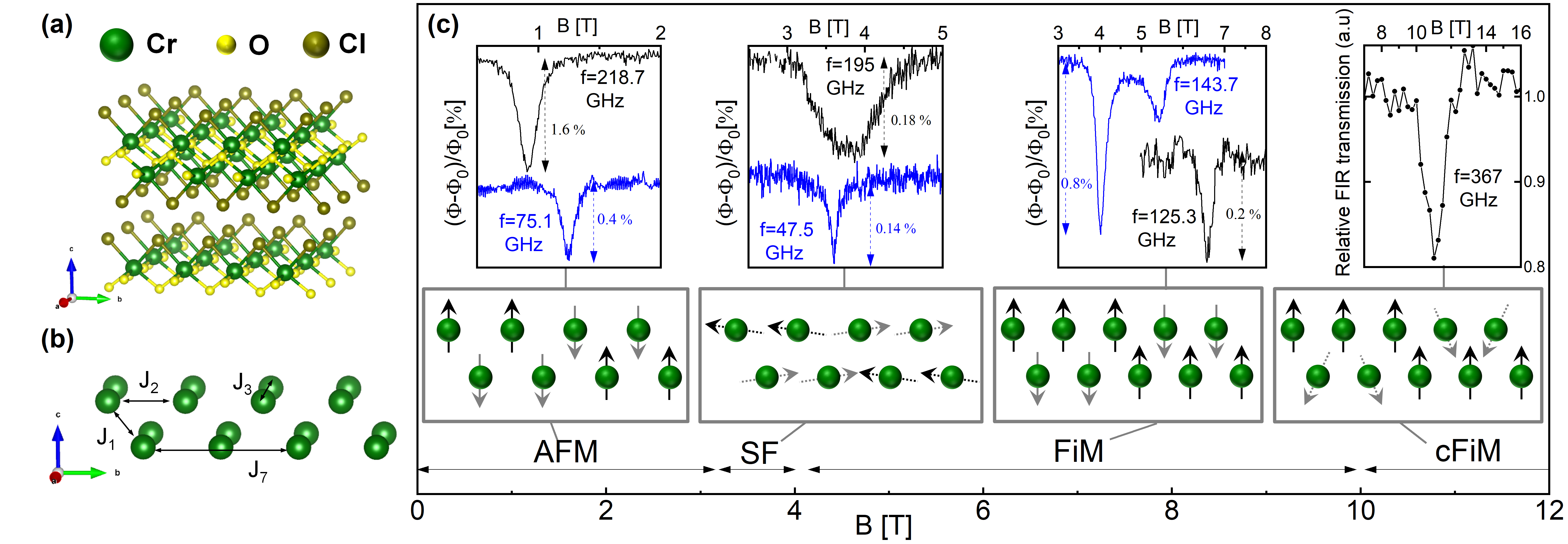}
\end{center}
\caption{(color online) (a) Crystal structure (2 layers, only half of the bottom layer is represented). (b) Definition of exchange coupling parameters between Cr spins. (c) Different phases reported in the literature (see text) with their schematic Cr spin configurations. The magnetic field boundary of each phase is approximately positioned based on previous experiments \cite{Reuvekamp2014,Gu2022,Pawbake2025} performed with upward magnetic fields sweeps at $T\sim 4$~K. Insets: representative raw data of magnetic field sweep-induced resonances at fixed microwave frequencies (specified to the first digit in GHz near each traces) in different magnetic phases. Relative changes of the photo-response phase $\phi$ are expressed with respect to their ``off-resonant'' value $\phi_{0}$. Data in black (blue) were taken with upward (downward) magnetic fields sweeps. The signal at $B\sim$11~T is obtained by Fourier Transform spectroscopy.}\label{Fig1}
\end{figure*}

The lattice of CrOCl consists of vdW layers made of two buckled planes of CrO terminated by Cl atoms, which at room temperature are stacked along the $c$-axis in an orthorhombic structure
with the Pmmn space group~\cite{Forsberg1962,Christensen1974} (see Fig.~\ref{Fig1}(a)). At low temperature, magnetic correlations become significant below $T_{mag}\sim27$~K, where an incommensurate spin density wave state is observed, leaving place to an AFM ground state with a monoclinic structure (space group P21/m~\cite{Angelkort2009,Reuvekamp2014}) below  the N\'eel temperature $T_{N}=13.5$~K. As the monoclinic angle $\alpha \simeq 90.06^{\circ}$~\cite{Angelkort2009} is very close to 90${^\circ}$, we will not distinguish between the $c$ and $c^{*}$ (perpendicular to the planes) directions in what follows. The leading exchange interactions parameters (defined in Fig.~\ref{Fig1}(b) and calculated in Ref.~\onlinecite{,Pawbake2025}) are $J_{1}$, $J_{2}$, $J_{3}$, and $J_{7}$. The large value of $J_{3}$ leads to ferromagnetically coupled stripes along the $a$-axis,  while 1D antiferromagnetic chains are observed along the $b$-axis, mainly due to the incompatible $J_{1}$ and $J_{7}$ interactions. A four-fold magnetic periodicity (illustrated in Fig.~\ref{Fig1}(c)) develops along the $b$-axis due to $J_{7}$ . Based on magnetic susceptibilities measurements~\cite{Angelkort2009}, CrOCl was considered to be a uniaxial antiferromagnet with the easy-axis being near the $c$-axis. More recent
DFT calculations of single-ion anisotropies~\cite{Gu2022,Zhang2023} confirmed the easiest $c$-axis and established the $a$-axis as an intermediate axis, and the $b$-axis as the hardest magnetization axis, but a complete mapping of the magnetic susceptibility in the $(c,a)$ and $(a,b)$ planes is lacking. Upon the application of a magnetic field along $c$-axis, different magnetic phases are successively observed: i) a canted phase with non-zero magnetization, previously attributed to a ``spin-flop'' (SF) transition~\cite{Angelkort2009,Gu2022}, ii) a ferrimagnetic phase with a five-fold magnetic periodicity~\cite{Gu2022} (as opposed to four-fold in the AFM and SF phase) recently confirmed by the observation of folded modes in Raman spectroscopy~\cite{Pawbake2025}. In higher magnetic fields, a succession of canted phases of different nature have been proposed~\cite{Reuvekamp2014,Pawbake2025} until the complete saturation of magnetization is observed, for a applied magnetic field of $B\sim30~T$. The sequence of the previously proposed field-induced magnetic phases is summarized in Fig.~\ref{Fig1}(c).

\begin{figure*}[]
\begin{center}
\includegraphics[width=18cm]{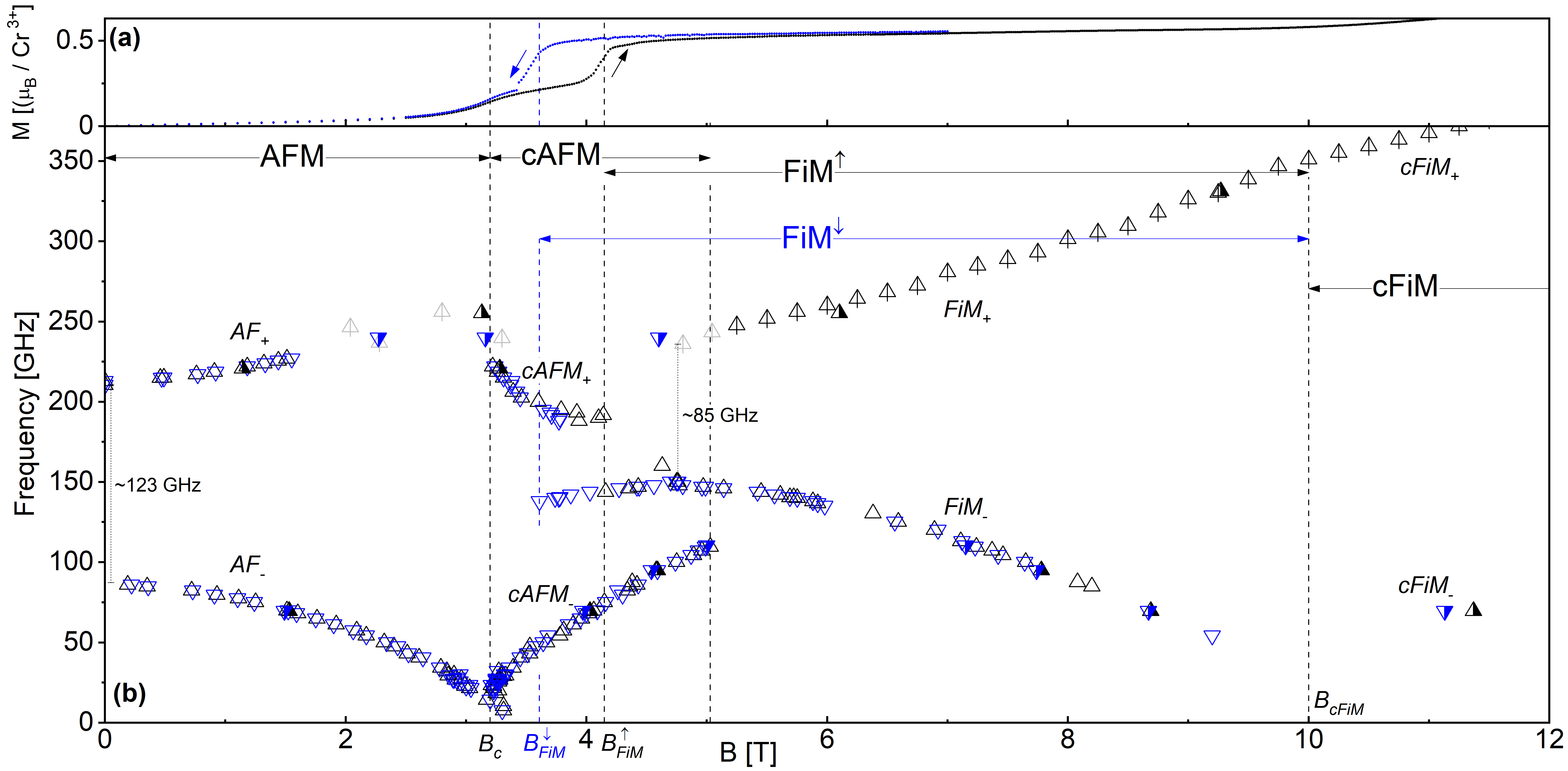}
\end{center}
\caption{(color online) (a) Magnetization traces for an upward (black) and downward (blue) magnetic field applied along the $c$-axis, at T$=4.2$K. (b) Magnetic resonances frequencies as a function of the magnetic field $B$ along the c-axis, for upward sweeps (open black up-triangles), or downward sweeps (open blue down-triangles). The sample temperature lies in the range $T = 3.9-4.8$~K depending on the microwave frequency. Triangles with a middle vertical dash were obtained at $T =4.2$~K by a FT spectrometer operating above $f\sim$ 250~GHz in the employed configuration. Additional data obtained with an EPR spectrometer at $T \sim5$~K are reported as half-filled triangles, with the same upward/downward sweeps convention. The location and extent of the different magnetic phases (AFM, cAFM, FiM, cFiM, see text) are indicated with double-sided horizontal arrows in black for upward sweeps, and blue for downward sweeps whenever hysteresis is observed.}\label{Fig2}
\end{figure*}

The bulk CrOCl crystals studied here were grown by a chemical vapor transport process described in Ref.~\onlinecite{Pawbake2025}. In the $f=[0-227$]~GHz frequency range,  microwave absorption experiments were performed  employing the phase-sensitive external resistive detection described in detail in Ref.~\onlinecite{Cho2023}, by monitoring the bolometric response of a resistive sensor placed below the sample.
Figure~\ref{Fig1}(c) shows typical examples of the sample's microwave absorption through the phase of the thermometer modulated photo-response. As one can see, absorption can be identified
by a clear single dip response in all the different magnetic-field-induced phases. At higher frequencies, Fourier-Transform (FT) far-infra-red (FIR) spectroscopy~\cite{LeMardele2024} was employed. Additional data were also obtained with an electron paramagnetic resonance (EPR) setup (see section II E in the supplemental material~\cite{SM}). All techniques probe the zero-wave number ($k=0$) excitations. In Fig.~\ref{Fig2}(b), we present the evolution of magnetic resonances when an external magnetic field up to 12~T is applied parallel to the $c$-axis. Magnetization traces measured on a sample issued from the same batch~\cite{Pawbake2025} (see also the supplemental material~\cite{SM}, section I), are shown in Fig.~\ref{Fig2}(a).

In the absence of magnetic field, two main absorptions are observed at frequencies of about $210.5\pm 1$ GHz and $87\pm 1$ GHz, which ressemble the zero-field magnon modes of an antiferromagnetic system with bi-axial anisotropy~\cite{Rezende2019}. These high and low energy modes, which we label $AF_{+}$ and $AF_{-}$, are split at $B=0$~T  by a rather large frequency gap of $(f_{AF_{+}}-f_{AF_{-}})\sim 123.5\pm 2$~GHz demonstrating a strong bi-axial (rather than uniaxial) anisotropy, and an intermediate axis in the $(a,b)$ plane. We note that no other magnon modes are observed in the antiferromagnetic phase up to the highest energies probed ($\sim900$~GHz). The evolution of the zero-field $AF_{-}$ magnon mode as a function of temperature is presented in section II D of the supplementary material~\cite{SM}.

As the magnetic field increases, the $AF_{+}$ ($AF_{-}$) disperses positively (negatively), until a sharp change is observed in the $3.2-3.26$~T region. Concomitantly (see Fig.~\ref{Fig2}(a)), the magnetization has departed from its null value in the AFM phase and the systems enters a new phase via what was previously attributed to a spin-flop transition \cite{Angelkort2009,Gu2022}. In this new phase, two magnon modes are observed. The low energy mode, $cAFM_{-}$, emerging from the $AF_{-}$ mode, and a high energy mode, $cAFM_{+}$, most likely emerging from the $AF_{+}$ mode. The $cAFM_{-}$ mode shows an initial steep increase from 14 to 27~GHz above $B_{c}=3.2$~T and then increases (sublinearly) with $B$, as usually observed in regimes where canted moments are progressively aligning toward the direction of the applied magnetic field. We note that the $cAFM_{-}$ magnon branch does not show any hysteresis in the vicinity of $B_{c}$. As discussed in Fig.~S7 in Ref.~\onlinecite{Gu2022}, the nature of this transition, first (spin-flop like) or second order, may be influenced by a non-purely uniaxial anisotropy, which is apparent in our AFM magnon spectra. Possible frustration emerging from longer range magnetic Heisenberg interactions may also play a role. The weak net magnetization values observed in our data and reported in the literature~\cite{Angelkort2009,Gu2022} would imply, in the case of a conventional spin-flop regime corresponding to all spins depicting a positive (i.e. along $B$) magnetization, a very large canting angle of the moments of these spins with respect to the $c$-axis (as pictured in Fig.~\ref{Fig1}(c)). This spin configuration is energetically unfavorable in the presence of single-ion anisotropy, which favors their alignment along the $c$-axis. On the other hand, thanks to the four-fold magnetic periodicity, other SF-like transitions can be envisioned, such as a transition where the down spins are only partially flipped (resulting in lower canted angles). Such alternative spin configurations are discussed in section II C in the supplementary material~\cite{SM}. For these reasons, we will in the remaining of this letter denominate this phase in a broader way as a ``canted AFM'' phase (cAFM) rather than a spin-flop regime.

The high energy mode ($cAFM_{+}$) frequency decreases with the magnetic field with a saturation or shallow minimum around $\sim3.86$~T, before disappearing above $\sim4.14$~T. In a biaxial AFM system undergoing a spin-flop transition, the upper magnon energy is expected to abruptly drop at the spin-flop magnetic field due to the magnetic moment reorientation, and asymptotically merge to a low field-independent value, which may not be necessarily null depending on in-plane anisotropies ~\cite{Rezende2019}. Therefore, the persistence of a resonance at frequencies similar to those of $AF_{+}$ for $B>B_{c}$ is not common, and reminiscent of the so-called ``orientation resonance'' discussed in Ref.~\onlinecite{Kobets2009}, which can be influenced by a coupling with the lower energy branch. Such a coupling may result from different effects. First, in the presence of an offset between the directions of $B$ and the symmetry axis (here, assumed to be the $c$-axis), the coupling between the two modes is allowed and the intersection between the branches
disappears~\cite{Kobets2009} (see other examples of magnon-magnon coupling observed for a magnetic field close to an intermediate axis~\cite{Velikov1970,Danshin1984,Cho2023}).
Additionally, in monoclinic AFMs where the Dzyaloshinskii-Moriya interaction (DMI) is allowed, ferromagnetic interactions can lead to independent invariants of second order in the magnetic energy,
making a magnetic moment induced along an easy axis deviate from that axis~\cite{Kobets2009,Velikov1970}, boosting the repulsion of the spin-flop modes.
In our case, $B$ is aligned within a few degrees to the $c$-axis, rendering the first effect small provided the easy-axis is indeed exactly along $c$. In summary, two parameters could help explaining more precisely the magnon modes behaviour in this region: the expected non-zero asymptotic energy of $cAFM_{+}$ (related, in principle, to the in-plane anisotropy), and a possible magnon-magnon coupling.

For an upward magnetic field above $\sim4$~T, a second rise is observed in the magnetization, previously attributed to collective magnetic moment flipping, triggering a transition to a FiM phase with a five-fold magnetic periodicity~\cite{Gu2022,Zhang2023,Pawbake2025}. In our data, a new resonance which we term $FiM_{-}$ appears at $B_{FiM}^{\uparrow}\sim4.15$~T. Its frequency displays a shallow maximum with increasing $B$ before decreasing in energy at fields above$\sim4.7$~T. At this magnetic field, magnetization approaches the plateau at 1/5 of the total magnetization attributed to the FiM phase. At higher energy, yet another new branch, $FiM_{+}$, is observed in the FIR and EPR-setup data at frequencies $f>236$~GHz, and increases in energy with $B$. It is likely that this branch also appears at $B_{FiM}^{\uparrow}$, but the limited resolution of FIR in this energy range prevents from discerning it. The $FiM_{+}$ and $FiM_{-}$ are split in frequency by $\sim 85$~GHz at $B=4.76$~T. A strong hysteresis in the magnon dispersion at the transition to/from the FiM  is observed and highlighted by the black (upward magnetic field) and blue (downward magnetic field) traces in Fig.~\ref{Fig2}(b). This is consistent with our magnetization data (Fig.~\ref{Fig2}(a)), with the ones in Refs.~\onlinecite{Angelkort2009,Gu2022}, and also with dielectric constant~\cite{Reuvekamp2014} or tunneling conductance ~\cite{Zhang2023} measurements. For downward sweeps, the $FiM_{-}$ branch persists down to $B_{FiM}^{down}\sim 3.6$~T, a magnetic field for which magnetization starts to ``leave'' the ferrimagnetic plateau value (we note that the magnetization value is similar at $B_{FiM}^{\uparrow}$ and $B_{FiM}^{\downarrow}$). Around the transition fields, for $B>B_{FiM}^{\uparrow}$ for upward sweeps or $B>B_{FiM}^{\downarrow}$ for downward sweeps, both the $cAFM_{-}$ and $FiM_{-}$ magnon branches are visible, suggesting coexisting FiM and cAFM phases. We note that while the lower $cAFM_{-}$ can coexist with $FiM_{-}$ both for upward and downward sweeps, it never is the case for $cAFM_{+}$, which disappears (appears) approximately when the $FiM_{-}$ branch appears (disappears) for upward (downward) magnetic field sweeps. This could be related to the fact that the existence of the $cAFM_{-}$ at such energies relies on DMI in the monoclinic structure (discussed above), which is replaced by an orthorhombic structure with 5-fold magnetic periodicity upon transition to the FiM phase~\cite{Gu2022,Pawbake2025}.

\begin{figure}[]
\begin{center}
\includegraphics[width= 8.5cm]{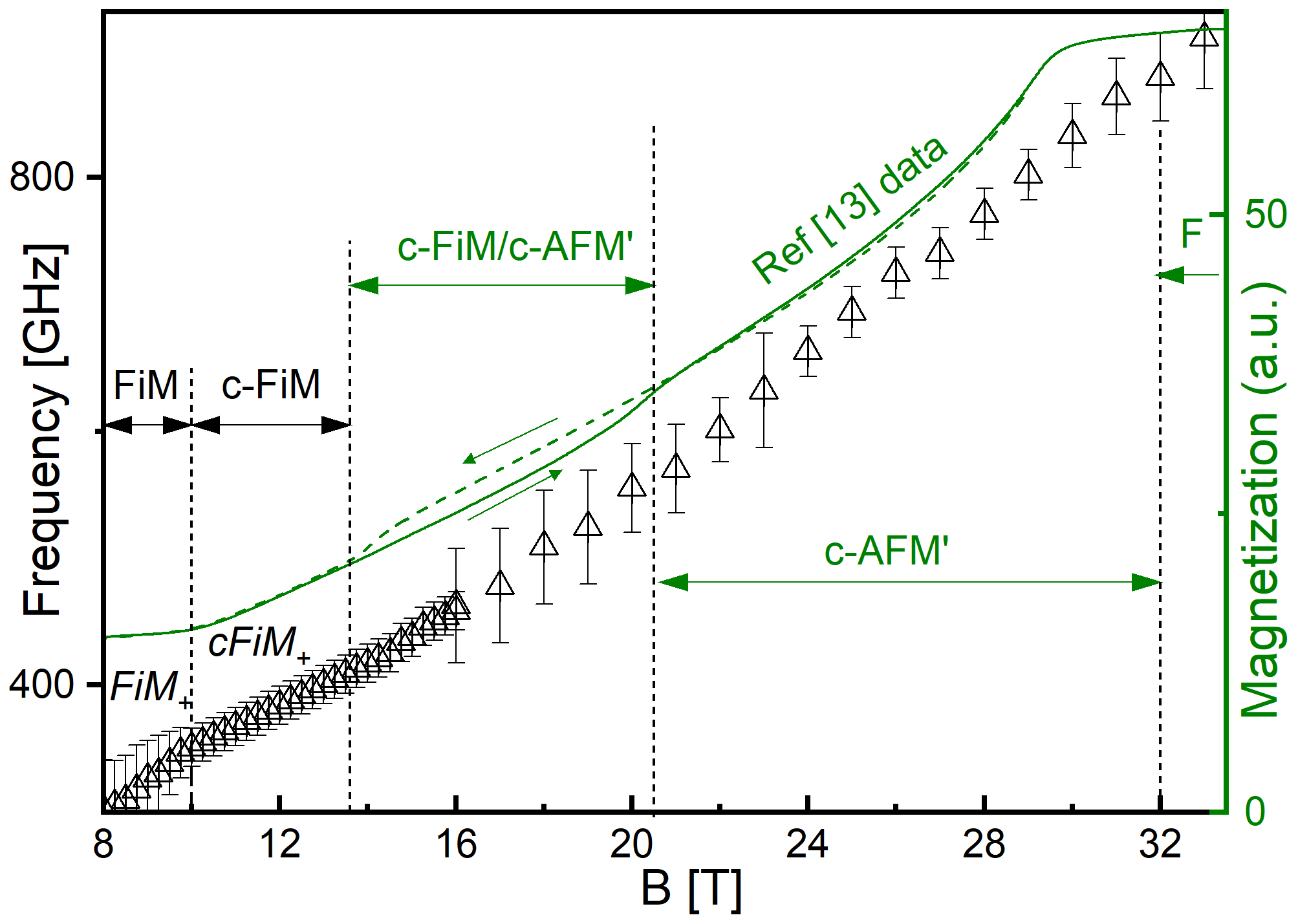}
\end{center}
\caption{(color online) Frequency of the main absorption observed in the FIR range, as a function of magnetic field (up triangles, left scale). Magnetization curves from Ref.~\onlinecite{Pawbake2025} (dark green solid (dashed) line for upward (downward) sweep, right scale).  Magnetic phases identified in Fig.~\ref{Fig2} (black text) and inferred from Ref.~\onlinecite{Pawbake2025} (dark green text).}\label{Fig4}
\end{figure}

Around $B_{cFiM}$=10~T, which corresponds to the magnetic field at which the magnetization departs from its FiM plateau value, the slope of the $FiM_{+}$ magnon's dispersion decreases. This presumably signals a phase change, from FiM to a canted FiM phase, indeed occurring above $B_{cFiM}$. A few resonance presumably corresponding to a second branch (termed $cFiM_{-}$ in Fig.~\ref{Fig2}(b)) can be observed at low energy in this field range (additional data are presented in the supplemental material~\cite{SM}, section II E). We note that the low frequency region $f<69.6$~GHz has not been probed for $B>B_{cFiM}$ in our experiments.
The new high energy branch, which we term $cFiM_{+}$, was followed for higher magnetic fields using FT-FIR spectroscopy (Fig.~\ref{Fig4}). Its frequency increases almost linearly from 10~T to $\sim13.6$~T, where another change in the slope (this time, an increase) is observed. This change matches the onset of an hysteretic cycle extending up to $\sim20.5$~T  (see the magnetization data of Ref.~\onlinecite{Pawbake2025} reported in Fig.~\ref{Fig4}) resulting from the competition of canted phases with different magnetic periodicites~\cite{Pawbake2025}. This branch increases approximately linearly from $\sim13.6$~T up to the highest studied magnetic field of $33$~T.

A full theoretical description of the complex experimental data presented here would require a Heisenberg Hamiltonian incorporating all the different frustrated exchange interactions (see Fig.~\ref{Fig1}(b)) and additional biaxial single-ion anisotropy terms. This is beyond the scope of the present letter and will be, together with a detailed discussion of the magnetization and low-energy magnetic excitations, the topic of a future communication.
Here, we present a simplified theoretical modelling of the low magnetic field magnon spectra, which will allow us to evaluate the relative strength of the effective exchange interactions and the anisotropy terms in the AFM phase when $B < B_{c}$. This model can be obtained from the general spin Hamiltonian proposed for a rhombic biaxial two-lattice AFM~\cite{Nagamiya1955}, which can be expressed in the following way~\cite{Cho2023}:

\begin{eqnarray}
\hat{\mathcal{H}} & = & \sum_{\langle ij\rangle}
J_{ij}\, \textbf{S}_{i} \cdot \textbf{S}_{j} + \sum_{i} \bigl[ -DS_{i}^{z2}+ E(S_{i}^{y2}-S_{i}^{x2})\bigr]
\nonumber \\
& - &  g \sum_i \mu_B B^\alpha S^\alpha \,,
\label{generalH}
\end{eqnarray}
where ${\bf S}_i$ are $S=3/2$ spins of Cr$^{3+}$ ions
and $g$ is the g-factor taken here to be isotropic with $g=$1.97 ~\cite{Angelkort2009}. The $x$, $y$ and $z$ cartesian coordinates are associated with the $a$, $b$ and $c$ crystallographic directions, respectively. $D$ and $E$ represent the $c$ easy-axis single ion anisotropy, and the difference between the $a$ intermediate and $b$ hard-axis single ion anisotropies, respectively. We encapsulate the effective exchange term resulting from the various (and possibly opposite) exchange contributions $J_{ij}$ in the term: $\hat{\mathcal{H}}_{\rm ex} = J_{eff} ({\bf S}_1\cdot {\bf S}_2)$. Analytical expressions of the AFM $k=0$ magnons magnetic field dispersion can then be formulated~\cite{Cho2023} and used to fit the experimental dispersion observed for $B<B_{c}$ in Fig.~\ref{Fig2}(b). A good fit of the experimental data at low magnetic fields (see the supplemental material~\cite{SM}, section II B) is obtained with the constraints: $J_{eff}D\sim ~5.9\pm0.6$~K$^{2}$ and $J_{eff}E\sim ~4.4\pm0.5$~K$^{2}$. The latter product can more simply be evaluated from the difference of the square energies of the $B=0$ magnon branches obtained from Ref.~\onlinecite{Cho2023}:  $ h^{2}(f_{AF_{+}}^{2}-f_{AF_{-}}^{2})=18 J_{eff} E $, which with the experimental $f_{AF_{+}}=210.5$~GHz and $f_{AF_{-}}=87$~GHz yields $J_{eff}E\sim ~4.7~$K$^{2}$. We note that the observation of a spin-flop transition (as opposed to an Ising transition) imposes $J_{eff}>2(D-E)$, corresponding to a weak anisotropy case. Together with the previous constraints, this imposes $(D-E)<0.87$~K, and $J_{eff}> 1.73$~K.

While a reasonable description of the low magnetic field ($B<B_{c}$) magnon spectra can be obtained with the former formalism, we note that the peculiar periodicities of the magnetic phases in CrOCl (4-fold periodic
AFM and cAFM phases, and 5-fold periodic FiM or cFiM phases) should a priori lead to a different magnon spectra, including higher energies branches, not observed experimentally.
The impact of the magnetic cell on the magnon spectra, potentially leading to a quantitative description of our data throughout the different magnetic field-induced phases, shall be the focus of further theoretical investigations.

To conclude, we have reported the observation of a multitude of magnon modes across the magnetic-field-induced phases of bulk CrOCl, a vdW AFM featuring antagonist spin interactions. Our data firmly establish the biaxial nature of CrOCl, evidence canted AFM modes reshaped by in-plane anisotropies and magnon coupling, and a hysteretic magnon spectrum with coexisting phases at the transition between a canted AFM and a FiM one. Competing exchange interactions allow for enhanced possibilities of tuning the magnons energies in a given material via the application of an external magnetic field.

We would like to thank K. Paillot, I. Breslavetz, D. Dufeu for technical assistance. L.A.V. and B.A.P. acknowledge support from ANR Grant
No. ANR-20-CE30-0015-01.  Z.S. was supported by project LUAUS25268 from Ministry of Education Youth and Sports (MEYS) and by the project Advanced Functional Nanorobots (reg. No. CZ.02.1.01/0.0/0.0/15 003/0000444 financed by the EFRR). Z.S. acknowledge the assistance provided by the Advanced Multiscale Materials for Key Enabling Technologies project, supported by the Ministry of Education, Youth, and Sports of the Czech Republic and Project No. CZ.02.01.01/00/22 008/0004558, Co-funded by the European Union. This work is also partially supported by ANR-23-QUAC-0004, CEFIPRA CSRP Project No. 7104-2, and by France 2030 government investment plan, managed by the French National Research Agency under Grant Reference PEPR SPIN–SPINMAT ANR-22-EXSP-0007.


\begin{thebibliography}{22}%
\makeatletter
\providecommand \@ifxundefined [1]{%
 \@ifx{#1\undefined}
}%
\providecommand \@ifnum [1]{%
 \ifnum #1\expandafter \@firstoftwo
 \else \expandafter \@secondoftwo
 \fi
}%
\providecommand \@ifx [1]{%
 \ifx #1\expandafter \@firstoftwo
 \else \expandafter \@secondoftwo
 \fi
}%
\providecommand \natexlab [1]{#1}%
\providecommand \enquote  [1]{``#1''}%
\providecommand \bibnamefont  [1]{#1}%
\providecommand \bibfnamefont [1]{#1}%
\providecommand \citenamefont [1]{#1}%
\providecommand \href@noop [0]{\@secondoftwo}%
\providecommand \href [0]{\begingroup \@sanitize@url \@href}%
\providecommand \@href[1]{\@@startlink{#1}\@@href}%
\providecommand \@@href[1]{\endgroup#1\@@endlink}%
\providecommand \@sanitize@url [0]{\catcode `\\12\catcode `\$12\catcode
  `\&12\catcode `\#12\catcode `\^12\catcode `\_12\catcode `\%12\relax}%
\providecommand \@@startlink[1]{}%
\providecommand \@@endlink[0]{}%
\providecommand \url  [0]{\begingroup\@sanitize@url \@url }%
\providecommand \@url [1]{\endgroup\@href {#1}{\urlprefix }}%
\providecommand \urlprefix  [0]{URL }%
\providecommand \Eprint [0]{\href }%
\providecommand \doibase [0]{https://doi.org/}%
\providecommand \selectlanguage [0]{\@gobble}%
\providecommand \bibinfo  [0]{\@secondoftwo}%
\providecommand \bibfield  [0]{\@secondoftwo}%
\providecommand \translation [1]{[#1]}%
\providecommand \BibitemOpen [0]{}%
\providecommand \bibitemStop [0]{}%
\providecommand \bibitemNoStop [0]{.\EOS\space}%
\providecommand \EOS [0]{\spacefactor3000\relax}%
\providecommand \BibitemShut  [1]{\csname bibitem#1\endcsname}%
\let\auto@bib@innerbib\@empty
\bibitem [{\citenamefont {Chumak}\ \emph {et~al.}(2015)\citenamefont {Chumak},
  \citenamefont {Vasyuchka}, \citenamefont {Serga},\ and\ \citenamefont
  {Hillebrands}}]{Chumak2015}%
  \BibitemOpen
  \bibfield  {author} {\bibinfo {author} {\bibfnamefont {A.~V.}\ \bibnamefont
  {Chumak}}, \bibinfo {author} {\bibfnamefont {V.}~\bibnamefont {Vasyuchka}},
  \bibinfo {author} {\bibfnamefont {A.}~\bibnamefont {Serga}},\ and\ \bibinfo
  {author} {\bibfnamefont {B.}~\bibnamefont {Hillebrands}},\ }\bibfield
  {title} {\bibinfo {title} {Magnon spintronics},\ }\href
  {https://doi.org/10.1038/nphys3347} {\bibfield  {journal} {\bibinfo
  {journal} {Nature Physics}\ }\textbf {\bibinfo {volume} {11}},\ \bibinfo
  {pages} {453} (\bibinfo {year} {2015})}\BibitemShut {NoStop}%
\bibitem [{\citenamefont {MacNeill}\ \emph {et~al.}(2019)\citenamefont
  {MacNeill}, \citenamefont {Hou}, \citenamefont {Klein}, \citenamefont
  {Zhang}, \citenamefont {Jarillo-Herrero},\ and\ \citenamefont
  {Liu}}]{MacNeill2019}%
  \BibitemOpen
  \bibfield  {author} {\bibinfo {author} {\bibfnamefont {D.}~\bibnamefont
  {MacNeill}}, \bibinfo {author} {\bibfnamefont {J.~T.}\ \bibnamefont {Hou}},
  \bibinfo {author} {\bibfnamefont {D.~R.}\ \bibnamefont {Klein}}, \bibinfo
  {author} {\bibfnamefont {P.}~\bibnamefont {Zhang}}, \bibinfo {author}
  {\bibfnamefont {P.}~\bibnamefont {Jarillo-Herrero}},\ and\ \bibinfo {author}
  {\bibfnamefont {L.}~\bibnamefont {Liu}},\ }\bibfield  {title} {\bibinfo
  {title} {Gigahertz frequency antiferromagnetic resonance and strong
  magnon-magnon coupling in the layered crystal ${\mathrm{crcl}}_{3}$},\ }\href
  {https://doi.org/10.1103/PhysRevLett.123.047204} {\bibfield  {journal}
  {\bibinfo  {journal} {Phys. Rev. Lett.}\ }\textbf {\bibinfo {volume} {123}},\
  \bibinfo {pages} {047204} (\bibinfo {year} {2019})}\BibitemShut {NoStop}%
\bibitem [{\citenamefont {Li}\ \emph {et~al.}(2023)\citenamefont {Li},
  \citenamefont {Dai}, \citenamefont {Ni}, \citenamefont {Zhang}, \citenamefont
  {Tang}, \citenamefont {Yang},\ and\ \citenamefont {Xu}}]{Li2023}%
  \BibitemOpen
  \bibfield  {author} {\bibinfo {author} {\bibfnamefont {W.}~\bibnamefont
  {Li}}, \bibinfo {author} {\bibfnamefont {Y.}~\bibnamefont {Dai}}, \bibinfo
  {author} {\bibfnamefont {L.}~\bibnamefont {Ni}}, \bibinfo {author}
  {\bibfnamefont {B.}~\bibnamefont {Zhang}}, \bibinfo {author} {\bibfnamefont
  {D.}~\bibnamefont {Tang}}, \bibinfo {author} {\bibfnamefont {Y.}~\bibnamefont
  {Yang}},\ and\ \bibinfo {author} {\bibfnamefont {Y.}~\bibnamefont {Xu}},\
  }\bibfield  {title} {\bibinfo {title} {Ultrastrong magnon magnon coupling and
  chirality switching in antiferromagnet crps4},\ }\href
  {https://doi.org/https://doi.org/10.1002/adfm.202303781} {\bibfield
  {journal} {\bibinfo  {journal} {Advanced Functional Materials}\ }\textbf
  {\bibinfo {volume} {33}},\ \bibinfo {pages} {2303781} (\bibinfo {year}
  {2023})}\BibitemShut {NoStop}%
\bibitem [{\citenamefont {Cho}\ \emph {et~al.}(2023)\citenamefont {Cho},
  \citenamefont {Pawbake}, \citenamefont {Aubergier}, \citenamefont {Barra},
  \citenamefont {Mosina}, \citenamefont {Sofer}, \citenamefont {Zhitomirsky},
  \citenamefont {Faugeras},\ and\ \citenamefont {Piot}}]{Cho2023}%
  \BibitemOpen
  \bibfield  {author} {\bibinfo {author} {\bibfnamefont {C.~W.}\ \bibnamefont
  {Cho}}, \bibinfo {author} {\bibfnamefont {A.}~\bibnamefont {Pawbake}},
  \bibinfo {author} {\bibfnamefont {N.}~\bibnamefont {Aubergier}}, \bibinfo
  {author} {\bibfnamefont {A.~L.}\ \bibnamefont {Barra}}, \bibinfo {author}
  {\bibfnamefont {K.}~\bibnamefont {Mosina}}, \bibinfo {author} {\bibfnamefont
  {Z.}~\bibnamefont {Sofer}}, \bibinfo {author} {\bibfnamefont {M.~E.}\
  \bibnamefont {Zhitomirsky}}, \bibinfo {author} {\bibfnamefont
  {C.}~\bibnamefont {Faugeras}},\ and\ \bibinfo {author} {\bibfnamefont
  {B.~A.}\ \bibnamefont {Piot}},\ }\bibfield  {title} {\bibinfo {title}
  {Microscopic parameters of the van der waals crsbr antiferromagnet from
  microwave absorption experiments},\ }\href
  {https://doi.org/10.1103/PhysRevB.107.094403} {\bibfield  {journal} {\bibinfo
   {journal} {Phys. Rev. B}\ }\textbf {\bibinfo {volume} {107}},\ \bibinfo
  {pages} {094403} (\bibinfo {year} {2023})}\BibitemShut {NoStop}%
\bibitem [{\citenamefont {Wosnitza}\ \emph {et~al.}(2016)\citenamefont
  {Wosnitza}, \citenamefont {Zvyagin},\ and\ \citenamefont
  {Zherlitsyn}}]{Wosnitza2016}%
  \BibitemOpen
  \bibfield  {author} {\bibinfo {author} {\bibfnamefont {J.}~\bibnamefont
  {Wosnitza}}, \bibinfo {author} {\bibfnamefont {S.~A.}\ \bibnamefont
  {Zvyagin}},\ and\ \bibinfo {author} {\bibfnamefont {S.}~\bibnamefont
  {Zherlitsyn}},\ }\bibfield  {title} {\bibinfo {title} {Frustrated magnets in
  high magnetic fields selected examples},\ }\href
  {https://doi.org/10.1088/0034-4885/79/7/074504} {\bibfield  {journal}
  {\bibinfo  {journal} {Reports on Progress in Physics}\ }\textbf {\bibinfo
  {volume} {79}},\ \bibinfo {pages} {074504} (\bibinfo {year}
  {2016})}\BibitemShut {NoStop}%
\bibitem [{\citenamefont {Thede}\ \emph {et~al.}(2014)\citenamefont {Thede},
  \citenamefont {Mannig}, \citenamefont {M\aa{}nsson}, \citenamefont
  {H\"uvonen}, \citenamefont {Khasanov}, \citenamefont {Morenzoni},\ and\
  \citenamefont {Zheludev}}]{Thede2014}%
  \BibitemOpen
  \bibfield  {author} {\bibinfo {author} {\bibfnamefont {M.}~\bibnamefont
  {Thede}}, \bibinfo {author} {\bibfnamefont {A.}~\bibnamefont {Mannig}},
  \bibinfo {author} {\bibfnamefont {M.}~\bibnamefont {M\aa{}nsson}}, \bibinfo
  {author} {\bibfnamefont {D.}~\bibnamefont {H\"uvonen}}, \bibinfo {author}
  {\bibfnamefont {R.}~\bibnamefont {Khasanov}}, \bibinfo {author}
  {\bibfnamefont {E.}~\bibnamefont {Morenzoni}},\ and\ \bibinfo {author}
  {\bibfnamefont {A.}~\bibnamefont {Zheludev}},\ }\bibfield  {title} {\bibinfo
  {title} {Pressure-induced quantum critical and multicritical points in a
  frustrated spin liquid},\ }\href
  {https://doi.org/10.1103/PhysRevLett.112.087204} {\bibfield  {journal}
  {\bibinfo  {journal} {Phys. Rev. Lett.}\ }\textbf {\bibinfo {volume} {112}},\
  \bibinfo {pages} {087204} (\bibinfo {year} {2014})}\BibitemShut {NoStop}%
\bibitem [{\citenamefont {Wang}\ \emph {et~al.}(2022)\citenamefont {Wang},
  \citenamefont {Bedoya-Pinto}, \citenamefont {Blei}, \citenamefont {Dismukes},
  \citenamefont {Hamo}, \citenamefont {Jenkins}, \citenamefont {Koperski},
  \citenamefont {Liu}, \citenamefont {Sun}, \citenamefont {Telford},
  \citenamefont {Kim}, \citenamefont {Augustin}, \citenamefont {Vool},
  \citenamefont {Yin}, \citenamefont {Li}, \citenamefont {Falin}, \citenamefont
  {Dean}, \citenamefont {Casanova}, \citenamefont {Evans}, \citenamefont
  {Chshiev}, \citenamefont {Mishchenko}, \citenamefont {Petrovic},
  \citenamefont {He}, \citenamefont {Zhao}, \citenamefont {Tsen}, \citenamefont
  {Gerardot}, \citenamefont {Brotons-Gisbert}, \citenamefont {Guguchia},
  \citenamefont {Roy}, \citenamefont {Tongay}, \citenamefont {Wang},
  \citenamefont {Hasan}, \citenamefont {Wrachtrup}, \citenamefont {Yacoby},
  \citenamefont {Fert}, \citenamefont {Parkin}, \citenamefont {Novoselov},
  \citenamefont {Dai}, \citenamefont {Balicas},\ and\ \citenamefont
  {Santos}}]{Wang2022}%
  \BibitemOpen
  \bibfield  {author} {\bibinfo {author} {\bibfnamefont {Q.~H.}\ \bibnamefont
  {Wang}}, \bibinfo {author} {\bibfnamefont {A.}~\bibnamefont {Bedoya-Pinto}},
  \bibinfo {author} {\bibfnamefont {M.}~\bibnamefont {Blei}}, \bibinfo {author}
  {\bibfnamefont {A.~H.}\ \bibnamefont {Dismukes}}, \bibinfo {author}
  {\bibfnamefont {A.}~\bibnamefont {Hamo}}, \bibinfo {author} {\bibfnamefont
  {S.}~\bibnamefont {Jenkins}}, \bibinfo {author} {\bibfnamefont
  {M.}~\bibnamefont {Koperski}}, \bibinfo {author} {\bibfnamefont
  {Y.}~\bibnamefont {Liu}}, \bibinfo {author} {\bibfnamefont {Q.-C.}\
  \bibnamefont {Sun}}, \bibinfo {author} {\bibfnamefont {E.~J.}\ \bibnamefont
  {Telford}}, \bibinfo {author} {\bibfnamefont {H.~H.}\ \bibnamefont {Kim}},
  \bibinfo {author} {\bibfnamefont {M.}~\bibnamefont {Augustin}}, \bibinfo
  {author} {\bibfnamefont {U.}~\bibnamefont {Vool}}, \bibinfo {author}
  {\bibfnamefont {J.-X.}\ \bibnamefont {Yin}}, \bibinfo {author} {\bibfnamefont
  {L.~H.}\ \bibnamefont {Li}}, \bibinfo {author} {\bibfnamefont
  {A.}~\bibnamefont {Falin}}, \bibinfo {author} {\bibfnamefont {C.~R.}\
  \bibnamefont {Dean}}, \bibinfo {author} {\bibfnamefont {F.}~\bibnamefont
  {Casanova}}, \bibinfo {author} {\bibfnamefont {R.~F.~L.}\ \bibnamefont
  {Evans}}, \bibinfo {author} {\bibfnamefont {M.}~\bibnamefont {Chshiev}},
  \bibinfo {author} {\bibfnamefont {A.}~\bibnamefont {Mishchenko}}, \bibinfo
  {author} {\bibfnamefont {C.}~\bibnamefont {Petrovic}}, \bibinfo {author}
  {\bibfnamefont {R.}~\bibnamefont {He}}, \bibinfo {author} {\bibfnamefont
  {L.}~\bibnamefont {Zhao}}, \bibinfo {author} {\bibfnamefont {A.~W.}\
  \bibnamefont {Tsen}}, \bibinfo {author} {\bibfnamefont {B.~D.}\ \bibnamefont
  {Gerardot}}, \bibinfo {author} {\bibfnamefont {M.}~\bibnamefont
  {Brotons-Gisbert}}, \bibinfo {author} {\bibfnamefont {Z.}~\bibnamefont
  {Guguchia}}, \bibinfo {author} {\bibfnamefont {X.}~\bibnamefont {Roy}},
  \bibinfo {author} {\bibfnamefont {S.}~\bibnamefont {Tongay}}, \bibinfo
  {author} {\bibfnamefont {Z.}~\bibnamefont {Wang}}, \bibinfo {author}
  {\bibfnamefont {M.~Z.}\ \bibnamefont {Hasan}}, \bibinfo {author}
  {\bibfnamefont {J.}~\bibnamefont {Wrachtrup}}, \bibinfo {author}
  {\bibfnamefont {A.}~\bibnamefont {Yacoby}}, \bibinfo {author} {\bibfnamefont
  {A.}~\bibnamefont {Fert}}, \bibinfo {author} {\bibfnamefont {S.}~\bibnamefont
  {Parkin}}, \bibinfo {author} {\bibfnamefont {K.~S.}\ \bibnamefont
  {Novoselov}}, \bibinfo {author} {\bibfnamefont {P.}~\bibnamefont {Dai}},
  \bibinfo {author} {\bibfnamefont {L.}~\bibnamefont {Balicas}},\ and\ \bibinfo
  {author} {\bibfnamefont {E.~J.~G.}\ \bibnamefont {Santos}},\ }\bibfield
  {title} {\bibinfo {title} {The magnetic genome of two-dimensional van der
  waals materials},\ }\href {https://doi.org/10.1021/acsnano.1c09150}
  {\bibfield  {journal} {\bibinfo  {journal} {ACS Nano}\ }\textbf {\bibinfo
  {volume} {16}},\ \bibinfo {pages} {6960} (\bibinfo {year} {2022})},\ \bibinfo
  {note} {pMID: 35442017}\BibitemShut {NoStop}%
\bibitem [{\citenamefont {Christensen}\ \emph {et~al.}(1974)\citenamefont
  {Christensen}, \citenamefont {Johansson},\ and\ \citenamefont
  {Quezel}}]{Christensen1974}%
  \BibitemOpen
  \bibfield  {author} {\bibinfo {author} {\bibfnamefont {A.~N.}\ \bibnamefont
  {Christensen}}, \bibinfo {author} {\bibfnamefont {T.}~\bibnamefont
  {Johansson}},\ and\ \bibinfo {author} {\bibfnamefont {S.}~\bibnamefont
  {Quezel}},\ }\href {https://doi.org/10.3891/acta.chem.scand.28a-1171}
  {\bibfield  {journal} {\bibinfo  {journal} {Acta Chemica Scandinavica}\
  }\textbf {\bibinfo {volume} {28a}},\ \bibinfo {pages} {1171} (\bibinfo {year}
  {1974})}\BibitemShut {NoStop}%
\bibitem [{\citenamefont {Gu}\ \emph {et~al.}(2022)\citenamefont {Gu},
  \citenamefont {Sun}, \citenamefont {Wang}, \citenamefont {Peng},
  \citenamefont {Zhu}, \citenamefont {Cheng}, \citenamefont {Yuan},
  \citenamefont {Lyu}, \citenamefont {Liu}, \citenamefont {Tan}, \citenamefont
  {Zhang}, \citenamefont {Gu}, \citenamefont {Wang}, \citenamefont {Wang},
  \citenamefont {Han}, \citenamefont {Watanabe}, \citenamefont {Taniguchi},
  \citenamefont {Yang}, \citenamefont {Zhang}, \citenamefont {Ji},
  \citenamefont {Tan},\ and\ \citenamefont {Ye}}]{Gu2022}%
  \BibitemOpen
  \bibfield  {author} {\bibinfo {author} {\bibfnamefont {P.}~\bibnamefont
  {Gu}}, \bibinfo {author} {\bibfnamefont {Y.}~\bibnamefont {Sun}}, \bibinfo
  {author} {\bibfnamefont {C.}~\bibnamefont {Wang}}, \bibinfo {author}
  {\bibfnamefont {Y.}~\bibnamefont {Peng}}, \bibinfo {author} {\bibfnamefont
  {Y.}~\bibnamefont {Zhu}}, \bibinfo {author} {\bibfnamefont {X.}~\bibnamefont
  {Cheng}}, \bibinfo {author} {\bibfnamefont {K.}~\bibnamefont {Yuan}},
  \bibinfo {author} {\bibfnamefont {C.}~\bibnamefont {Lyu}}, \bibinfo {author}
  {\bibfnamefont {X.}~\bibnamefont {Liu}}, \bibinfo {author} {\bibfnamefont
  {Q.}~\bibnamefont {Tan}}, \bibinfo {author} {\bibfnamefont {Q.}~\bibnamefont
  {Zhang}}, \bibinfo {author} {\bibfnamefont {L.}~\bibnamefont {Gu}}, \bibinfo
  {author} {\bibfnamefont {Z.}~\bibnamefont {Wang}}, \bibinfo {author}
  {\bibfnamefont {H.}~\bibnamefont {Wang}}, \bibinfo {author} {\bibfnamefont
  {Z.}~\bibnamefont {Han}}, \bibinfo {author} {\bibfnamefont {K.}~\bibnamefont
  {Watanabe}}, \bibinfo {author} {\bibfnamefont {T.}~\bibnamefont {Taniguchi}},
  \bibinfo {author} {\bibfnamefont {J.}~\bibnamefont {Yang}}, \bibinfo {author}
  {\bibfnamefont {J.}~\bibnamefont {Zhang}}, \bibinfo {author} {\bibfnamefont
  {W.}~\bibnamefont {Ji}}, \bibinfo {author} {\bibfnamefont {P.-H.}\
  \bibnamefont {Tan}},\ and\ \bibinfo {author} {\bibfnamefont {Y.}~\bibnamefont
  {Ye}},\ }\bibfield  {title} {\bibinfo {title} {Magnetic phase transitions and
  magnetoelastic coupling in a two-dimensional stripy antiferromagnet},\ }\href
  {https://doi.org/10.1021/acs.nanolett.1c04373} {\bibfield  {journal}
  {\bibinfo  {journal} {Nano Letters}\ }\textbf {\bibinfo {volume} {22}},\
  \bibinfo {pages} {1233} (\bibinfo {year} {2022})},\ \bibinfo {note} {pMID:
  35041438},\ \Eprint
  {https://arxiv.org/abs/https://doi.org/10.1021/acs.nanolett.1c04373}
  {https://doi.org/10.1021/acs.nanolett.1c04373} \BibitemShut {NoStop}%
\bibitem [{\citenamefont {Zhang}\ \emph {et~al.}(2023)\citenamefont {Zhang},
  \citenamefont {Hu}, \citenamefont {Huang}, \citenamefont {Hua}, \citenamefont
  {Cheng}, \citenamefont {Liu}, \citenamefont {Song}, \citenamefont {Wang},
  \citenamefont {Lu}, \citenamefont {He}, \citenamefont {Cao}, \citenamefont
  {Xu}, \citenamefont {Lu}, \citenamefont {Yang},\ and\ \citenamefont
  {Zheng}}]{Zhang2023}%
  \BibitemOpen
  \bibfield  {author} {\bibinfo {author} {\bibfnamefont {M.}~\bibnamefont
  {Zhang}}, \bibinfo {author} {\bibfnamefont {Q.}~\bibnamefont {Hu}}, \bibinfo
  {author} {\bibfnamefont {Y.}~\bibnamefont {Huang}}, \bibinfo {author}
  {\bibfnamefont {C.}~\bibnamefont {Hua}}, \bibinfo {author} {\bibfnamefont
  {M.}~\bibnamefont {Cheng}}, \bibinfo {author} {\bibfnamefont
  {Z.}~\bibnamefont {Liu}}, \bibinfo {author} {\bibfnamefont {S.}~\bibnamefont
  {Song}}, \bibinfo {author} {\bibfnamefont {F.}~\bibnamefont {Wang}}, \bibinfo
  {author} {\bibfnamefont {H.}~\bibnamefont {Lu}}, \bibinfo {author}
  {\bibfnamefont {P.}~\bibnamefont {He}}, \bibinfo {author} {\bibfnamefont
  {G.-H.}\ \bibnamefont {Cao}}, \bibinfo {author} {\bibfnamefont {Z.-A.}\
  \bibnamefont {Xu}}, \bibinfo {author} {\bibfnamefont {Y.}~\bibnamefont {Lu}},
  \bibinfo {author} {\bibfnamefont {J.}~\bibnamefont {Yang}},\ and\ \bibinfo
  {author} {\bibfnamefont {Y.}~\bibnamefont {Zheng}},\ }\bibfield  {title}
  {\bibinfo {title} {Spin-lattice coupled metamagnetism in frustrated van der
  waals magnet crocl},\ }\href
  {https://doi.org/https://doi.org/10.1002/smll.202300964} {\bibfield
  {journal} {\bibinfo  {journal} {Small}\ }\textbf {\bibinfo {volume} {19}},\
  \bibinfo {pages} {2300964} (\bibinfo {year} {2023})}\BibitemShut {NoStop}%
\bibitem [{\citenamefont {Gong}\ \emph {et~al.}(2017)\citenamefont {Gong},
  \citenamefont {Li}, \citenamefont {Li}, \citenamefont {Ji}, \citenamefont
  {Stern}, \citenamefont {Xia}, \citenamefont {Cao}, \citenamefont {Bao},
  \citenamefont {Wang}, \citenamefont {Wang}, \citenamefont {Qiu},
  \citenamefont {Cava}, \citenamefont {Louie}, \citenamefont {Xia},\ and\
  \citenamefont {Zhang}}]{Gong2017}%
  \BibitemOpen
  \bibfield  {author} {\bibinfo {author} {\bibfnamefont {C.}~\bibnamefont
  {Gong}}, \bibinfo {author} {\bibfnamefont {L.}~\bibnamefont {Li}}, \bibinfo
  {author} {\bibfnamefont {Z.}~\bibnamefont {Li}}, \bibinfo {author}
  {\bibfnamefont {H.}~\bibnamefont {Ji}}, \bibinfo {author} {\bibfnamefont
  {A.}~\bibnamefont {Stern}}, \bibinfo {author} {\bibfnamefont
  {Y.}~\bibnamefont {Xia}}, \bibinfo {author} {\bibfnamefont {T.}~\bibnamefont
  {Cao}}, \bibinfo {author} {\bibfnamefont {W.}~\bibnamefont {Bao}}, \bibinfo
  {author} {\bibfnamefont {C.}~\bibnamefont {Wang}}, \bibinfo {author}
  {\bibfnamefont {Y.}~\bibnamefont {Wang}}, \bibinfo {author} {\bibfnamefont
  {Z.~Q.}\ \bibnamefont {Qiu}}, \bibinfo {author} {\bibfnamefont {R.~J.}\
  \bibnamefont {Cava}}, \bibinfo {author} {\bibfnamefont {S.~G.}\ \bibnamefont
  {Louie}}, \bibinfo {author} {\bibfnamefont {J.}~\bibnamefont {Xia}},\ and\
  \bibinfo {author} {\bibfnamefont {X.}~\bibnamefont {Zhang}},\ }\bibfield
  {title} {\bibinfo {title} {Discovery of intrinsic ferromagnetism in
  two-dimensional van der waals crystals},\ }\href
  {https://doi.org/10.1038/nature22060} {\bibfield  {journal} {\bibinfo
  {journal} {Nature}\ }\textbf {\bibinfo {volume} {546}},\ \bibinfo {pages}
  {265} (\bibinfo {year} {2017})}\BibitemShut {NoStop}%
\bibitem [{\citenamefont {Reuvekamp}(2014)}]{Reuvekamp2014}%
  \BibitemOpen
  \bibfield  {author} {\bibinfo {author} {\bibfnamefont {P.}~\bibnamefont
  {Reuvekamp}},\ }\emph {\bibinfo {title} {Investigation into the magnetic and
  the structural properties of two low-dimensional antiferromagnets TiPO4 and
  CrOCl}},\ \href@noop {} {Ph.D. thesis},\ \bibinfo  {school} {University of
  Stuttgart: Stuttgart, Germany} (\bibinfo {year} {2014})\BibitemShut {NoStop}%
\bibitem [{\citenamefont {Pawbake}\ \emph {et~al.}(2025)\citenamefont
  {Pawbake}, \citenamefont {Petot}, \citenamefont {Le~Mardel\'e}, \citenamefont
  {Riccardi}, \citenamefont {L\'ev\^eque}, \citenamefont {Piot}, \citenamefont
  {Orlita}, \citenamefont {Coraux}, \citenamefont {Hubert}, \citenamefont
  {Dzian}, \citenamefont {Veis}, \citenamefont {Skourski}, \citenamefont {Wu},
  \citenamefont {Sofer}, \citenamefont {Gr\'emaud}, \citenamefont {Sa\'ul},\
  and\ \citenamefont {Faugeras}}]{Pawbake2025}%
  \BibitemOpen
  \bibfield  {author} {\bibinfo {author} {\bibfnamefont {A.}~\bibnamefont
  {Pawbake}}, \bibinfo {author} {\bibfnamefont {F.}~\bibnamefont {Petot}},
  \bibinfo {author} {\bibfnamefont {F.}~\bibnamefont {Le~Mardel\'e}}, \bibinfo
  {author} {\bibfnamefont {T.}~\bibnamefont {Riccardi}}, \bibinfo {author}
  {\bibfnamefont {J.}~\bibnamefont {L\'ev\^eque}}, \bibinfo {author}
  {\bibfnamefont {B.~A.}\ \bibnamefont {Piot}}, \bibinfo {author}
  {\bibfnamefont {M.}~\bibnamefont {Orlita}}, \bibinfo {author} {\bibfnamefont
  {J.}~\bibnamefont {Coraux}}, \bibinfo {author} {\bibfnamefont
  {M.}~\bibnamefont {Hubert}}, \bibinfo {author} {\bibfnamefont
  {J.}~\bibnamefont {Dzian}}, \bibinfo {author} {\bibfnamefont
  {M.}~\bibnamefont {Veis}}, \bibinfo {author} {\bibfnamefont {Y.}~\bibnamefont
  {Skourski}}, \bibinfo {author} {\bibfnamefont {B.}~\bibnamefont {Wu}},
  \bibinfo {author} {\bibfnamefont {Z.}~\bibnamefont {Sofer}}, \bibinfo
  {author} {\bibfnamefont {B.}~\bibnamefont {Gr\'emaud}}, \bibinfo {author}
  {\bibfnamefont {A.}~\bibnamefont {Sa\'ul}},\ and\ \bibinfo {author}
  {\bibfnamefont {C.}~\bibnamefont {Faugeras}},\ }\bibfield  {title} {\bibinfo
  {title} {Magnetic phases and zone-folded phonons in a frustrated van der
  waals magnet},\ }\href {https://doi.org/10.1021/acsnano.5c03174} {\bibfield
  {journal} {\bibinfo  {journal} {ACS Nano}\ }\textbf {\bibinfo {volume}
  {19}},\ \bibinfo {pages} {23693} (\bibinfo {year} {2025})}\BibitemShut
  {NoStop}%
\bibitem [{\citenamefont {Forsberg}(1962)}]{Forsberg1962}%
  \BibitemOpen
  \bibfield  {author} {\bibinfo {author} {\bibfnamefont {H.-E.}\ \bibnamefont
  {Forsberg}},\ }\href {https://doi.org/10.3891/acta.chem.scand.16-0777}
  {\bibfield  {journal} {\bibinfo  {journal} {Acta Chemica Scandinavica}\
  }\textbf {\bibinfo {volume} {16}},\ \bibinfo {pages} {777} (\bibinfo {year}
  {1962})}\BibitemShut {NoStop}%
\bibitem [{\citenamefont {Angelkort}\ \emph {et~al.}(2009)\citenamefont
  {Angelkort}, \citenamefont {W\"olfel}, \citenamefont {Sch\"onleber},
  \citenamefont {van Smaalen},\ and\ \citenamefont {Kremer}}]{Angelkort2009}%
  \BibitemOpen
  \bibfield  {author} {\bibinfo {author} {\bibfnamefont {J.}~\bibnamefont
  {Angelkort}}, \bibinfo {author} {\bibfnamefont {A.}~\bibnamefont {W\"olfel}},
  \bibinfo {author} {\bibfnamefont {A.}~\bibnamefont {Sch\"onleber}}, \bibinfo
  {author} {\bibfnamefont {S.}~\bibnamefont {van Smaalen}},\ and\ \bibinfo
  {author} {\bibfnamefont {R.~K.}\ \bibnamefont {Kremer}},\ }\bibfield  {title}
  {\bibinfo {title} {Observation of strong magnetoelastic coupling in a
  first-order phase transition of crocl},\ }\href
  {https://doi.org/10.1103/PhysRevB.80.144416} {\bibfield  {journal} {\bibinfo
  {journal} {Phys. Rev. B}\ }\textbf {\bibinfo {volume} {80}},\ \bibinfo
  {pages} {144416} (\bibinfo {year} {2009})}\BibitemShut {NoStop}%
\bibitem [{\citenamefont {Le~Mardel\'e}\ \emph {et~al.}(2024)\citenamefont
  {Le~Mardel\'e}, \citenamefont {El~Mendili}, \citenamefont {Zhitomirsky},
  \citenamefont {Mohelsky}, \citenamefont {Jana}, \citenamefont {Plutnarova},
  \citenamefont {Sofer}, \citenamefont {Faugeras}, \citenamefont {Potemski},\
  and\ \citenamefont {Orlita}}]{LeMardele2024}%
  \BibitemOpen
  \bibfield  {author} {\bibinfo {author} {\bibfnamefont {F.}~\bibnamefont
  {Le~Mardel\'e}}, \bibinfo {author} {\bibfnamefont {A.}~\bibnamefont
  {El~Mendili}}, \bibinfo {author} {\bibfnamefont {M.~E.}\ \bibnamefont
  {Zhitomirsky}}, \bibinfo {author} {\bibfnamefont {I.}~\bibnamefont
  {Mohelsky}}, \bibinfo {author} {\bibfnamefont {D.}~\bibnamefont {Jana}},
  \bibinfo {author} {\bibfnamefont {I.}~\bibnamefont {Plutnarova}}, \bibinfo
  {author} {\bibfnamefont {Z.}~\bibnamefont {Sofer}}, \bibinfo {author}
  {\bibfnamefont {C.}~\bibnamefont {Faugeras}}, \bibinfo {author}
  {\bibfnamefont {M.}~\bibnamefont {Potemski}},\ and\ \bibinfo {author}
  {\bibfnamefont {M.}~\bibnamefont {Orlita}},\ }\bibfield  {title} {\bibinfo
  {title} {Transverse and longitudinal magnons in the strongly anisotropic
  antiferromagnet ${\mathrm{fepse}}_{3}$},\ }\href
  {https://doi.org/10.1103/PhysRevB.109.134410} {\bibfield  {journal} {\bibinfo
   {journal} {Phys. Rev. B}\ }\textbf {\bibinfo {volume} {109}},\ \bibinfo
  {pages} {134410} (\bibinfo {year} {2024})}\BibitemShut {NoStop}%
\bibitem [{SM()}]{SM}%
  \BibitemOpen
  \bibinfo {note} {See the supplemental material at (URL will be inserted by
  publisher)}\BibitemShut {NoStop}%
\bibitem [{\citenamefont {Rezende}\ \emph {et~al.}(2019)\citenamefont
  {Rezende}, \citenamefont {Azevedo},\ and\ \citenamefont
  {Rodriguez-Suarez}}]{Rezende2019}%
  \BibitemOpen
  \bibfield  {author} {\bibinfo {author} {\bibfnamefont {S.~M.}\ \bibnamefont
  {Rezende}}, \bibinfo {author} {\bibfnamefont {A.}~\bibnamefont {Azevedo}},\
  and\ \bibinfo {author} {\bibfnamefont {R.~L.}\ \bibnamefont
  {Rodriguez-Suarez}},\ }\bibfield  {title} {\bibinfo {title} {Introduction to
  antiferromagnetic magnons},\ }\href {https://doi.org/10.1063/1.5109132}
  {\bibfield  {journal} {\bibinfo  {journal} {Journal of Applied Physics}\
  }\textbf {\bibinfo {volume} {126}},\ \bibinfo {pages} {151101} (\bibinfo
  {year} {2019})}\BibitemShut {NoStop}%
\bibitem [{\citenamefont {Kobets}\ \emph {et~al.}(2009)\citenamefont {Kobets},
  \citenamefont {Dergachev}, \citenamefont {Gnatchenko}, \citenamefont
  {Khatsko}, \citenamefont {Vysochanskii},\ and\ \citenamefont
  {Gurzan}}]{Kobets2009}%
  \BibitemOpen
  \bibfield  {author} {\bibinfo {author} {\bibfnamefont {M.~I.}\ \bibnamefont
  {Kobets}}, \bibinfo {author} {\bibfnamefont {K.~G.}\ \bibnamefont
  {Dergachev}}, \bibinfo {author} {\bibfnamefont {S.~L.}\ \bibnamefont
  {Gnatchenko}}, \bibinfo {author} {\bibfnamefont {E.~N.}\ \bibnamefont
  {Khatsko}}, \bibinfo {author} {\bibfnamefont {Y.~M.}\ \bibnamefont
  {Vysochanskii}},\ and\ \bibinfo {author} {\bibfnamefont {M.~I.}\ \bibnamefont
  {Gurzan}},\ }\bibfield  {title} {\bibinfo {title} {Antiferromagnetic
  resonance in mn2p2s6},\ }\href {https://doi.org/10.1063/1.3272560} {\bibfield
   {journal} {\bibinfo  {journal} {Low Temperature Physics}\ }\textbf {\bibinfo
  {volume} {35}},\ \bibinfo {pages} {930} (\bibinfo {year} {2009})}\BibitemShut
  {NoStop}%
\bibitem [{\citenamefont {Velikov}\ \emph {et~al.}(1970)\citenamefont
  {Velikov}, \citenamefont {Mironov},\ and\ \citenamefont
  {Rudashevskii}}]{Velikov1970}%
  \BibitemOpen
  \bibfield  {author} {\bibinfo {author} {\bibfnamefont {L.~V.}\ \bibnamefont
  {Velikov}}, \bibinfo {author} {\bibfnamefont {S.~V.}\ \bibnamefont
  {Mironov}},\ and\ \bibinfo {author} {\bibfnamefont {E.~G.}\ \bibnamefont
  {Rudashevskii}},\ }\href@noop {} {\bibfield  {journal} {\bibinfo  {journal}
  {Sov. Phys. JETP}\ }\textbf {\bibinfo {volume} {30}},\ \bibinfo {pages} {428}
  (\bibinfo {year} {1970})},\ \bibinfo {note} {zh. Eksp. Teor. Fiz. 57, 781
  1969}\BibitemShut {NoStop}%
\bibitem [{\citenamefont {Danshin}\ \emph {et~al.}(1984)\citenamefont
  {Danshin}, \citenamefont {Zeltser},\ and\ \citenamefont
  {Popov}}]{Danshin1984}%
  \BibitemOpen
  \bibfield  {author} {\bibinfo {author} {\bibfnamefont {N.~K.}\ \bibnamefont
  {Danshin}}, \bibinfo {author} {\bibfnamefont {A.~S.}\ \bibnamefont
  {Zeltser}},\ and\ \bibinfo {author} {\bibfnamefont {V.~A.}\ \bibnamefont
  {Popov}},\ }\bibfield  {title} {\bibinfo {title} {Excitation anisotropy of
  coupled spin-vibration branches in antiferromagnetic cuci2-2h2o},\ }\href
  {https://doi.org/10.1063/10.0031205} {\bibfield  {journal} {\bibinfo
  {journal} {Soviet Journal Low Temperature Physics}\ }\textbf {\bibinfo
  {volume} {10}},\ \bibinfo {pages} {613} (\bibinfo {year} {1984})}\BibitemShut
  {NoStop}%
\bibitem [{\citenamefont {Nagamiya}\ \emph {et~al.}(1955)\citenamefont
  {Nagamiya}, \citenamefont {Yosida},\ and\ \citenamefont
  {Kubo}}]{Nagamiya1955}%
  \BibitemOpen
  \bibfield  {author} {\bibinfo {author} {\bibfnamefont {T.}~\bibnamefont
  {Nagamiya}}, \bibinfo {author} {\bibfnamefont {K.}~\bibnamefont {Yosida}},\
  and\ \bibinfo {author} {\bibfnamefont {R.}~\bibnamefont {Kubo}},\ }\bibfield
  {title} {\bibinfo {title} {Antiferromagnetism},\ }\href
  {https://doi.org/10.1080/00018735500101154} {\bibfield  {journal} {\bibinfo
  {journal} {Advances in Physics}\ }\textbf {\bibinfo {volume} {4}},\ \bibinfo
  {pages} {1} (\bibinfo {year} {1955})},\ \Eprint
  {https://arxiv.org/abs/https://doi.org/10.1080/00018735500101154}
  {https://doi.org/10.1080/00018735500101154} \BibitemShut {NoStop}%
\end{thebibliography}
\end{document}